\documentclass[prd,twocolumn]{revtex4}
\usepackage{bm}
\usepackage{epsfig}
\usepackage{amsmath}
\newcommand{\be}{\begin{equation}}
\newcommand{\ee}{\end{equation}}
\newcommand{\ba}{\begin{eqnarray}}
\newcommand{\ea}{\end{eqnarray}}
\newcommand{\bes}{\begin{subequations}}
\newcommand{\ees}{\end{subequations}}

\newcommand{\bi}{\begin{itemize}}
\newcommand{\ei}{\end{itemize}}

\begin{document}
\title{Electroweak-scale mirror fermions, $\mu \rightarrow e\,\gamma$
and $\tau \rightarrow \mu\,\gamma$}
\author{P.Q. Hung}
\email[]{pqh@virginia.edu}
\affiliation{Dept. of Physics, University of Virginia, \\
382 McCormick Road, P. O. Box 400714, Charlottesville, Virginia 22904-4714, 
USA}
\date{\today}
\begin{abstract}
The Lepton Flavour Violating (LFV) processes $\mu \rightarrow e\,\gamma$ and 
$\tau \rightarrow \mu\,\gamma$ are estimated in a model of
electroweak-scale right-handed neutrinos. The present bounds on
the branching ratios, $B(\mu \rightarrow e\,\gamma)<
1.2 \times 10^{-11}$, $B(\tau \rightarrow \mu\,\gamma) < 6.8 \times 10^{-8}$ (BaBar) and
$< 4.5 \times 10^{-8}$ (Belle), 
puts strong constraints on the parameters
of the model. This constraint links low energy rare decay processes
to high-energy phenomena (e.g. decay lengths
of the mirror charged leptons which are important in the search for
the telltale like-sign dilepton events present in the
model of electroweak-scale right-handed neutrinos). The model
can be tested at future colliders (LHC, ILC) and at MEG and/or
B factories.
\end{abstract}
\maketitle

\section{Introduction}

The nature and origin of neutrino masses are undoubtedly among the most
important questions in particle physics, experimentally and theoretically.
Neutrino oscillation experiments and astrophysical arguments indicate that
neutrinos that interact with normal matter at tree level are very light
compared with all known charged particles, typically with masses less
than an electronvolt. 
There exists plausible models of light neutrino masses which
could, in principle, be tested experimentally. One of such models is the
famous see-saw mechanism  \cite{seesaw} where a lepton-number conserving 
($\Delta L=0$) Dirac neutrino mass term, $m_D\,\bar{\nu}_L \nu_R$,
is combined with a lepton-number violating ($\Delta L=2$) Majorana
mass term, $M_R \nu_R^{T} \sigma_2 \nu_R$ where it is usually assumed
that $M_R \gg m_D$, to yield a ``tiny'' mass eigenvalue $-m_D^2/M_R$ and
a ``very large'' one $M_R$. Since in a generic framework, the only 
``knowledge'' that one has at the present time is the smallness of the
ratio $-m_D^2/M_R$, the question of how small $m_D$ is and how
large $M_R$ could be is rather model-dependent. The most popular scenario 
is one in which
$m_D$ is related to the electroweak symmetry breaking scale while
$M_R$ is in general related to some Grand Unified scale. It is in
particular the relationship of the Majorana scale $M_R$ to some new physics
that is of great interest since its probe would reveal not only
the see-saw mechanism but also what type of new physics one might be
dealing with.

A very interesting connection between $M_R$ and the scale above which
parity is restored was made by \cite{rabigoran}
in which the Standard Model (SM) is extended to $SU(2)_L \otimes
SU(2)_R \otimes U(1)_{B-L}$. In this model, the
finiteness, albeit small, of the light neutrino mass $m_{\nu}$ is related to
the finiteness of the $SU(2)_R$ gauge boson masses $M_{W_R} \gg M_{W_L}$ with
$m_{\nu}$ vanishing in the limit $M_{W_R} \rightarrow \infty$. The V-A nature
of the weak interactions of the SM fermions is recovered in this limit.
It is extremely intriguing that parity restoration is linked in the
left-right (LR) symmetric model to the non-vanishing value of the neutrino mass. 
The strength of V+A interactions vanishes in the limit of zero neutrino masses.

Can parity restoration be accomplished within the SM gauge sector? 
This issue was addressed in \cite{hung1,hung2} where the gauge group
is simply the SM $SU(2) \otimes U(1)$. (Notice that the subscripts $L$ and
$Y$ are deliberately omitted for reasons to be given below.) However,
one now has, for every SM left-handed doublet such as e.g.
$(\nu_e, e)_L$, a {\em heavy} mirror {\em right-handed}
doublet e.g. $(\nu^{M}_e,e^{M})_R$ \cite{mirror}. Similarly, for every SM singlet
such as e.g. $e_R$, one has a {\em heavy} {\em left-handed} mirror singlet
$e^{M}_L$. (The content for the quarks and their mirror
counterparts are listed in \cite{hung1}.) One word of caution
is in order here. What we mean by ``mirror'' is simply the
aforementioned assignments for fermions and nothing else, e.g.
there is {\em no} mirror gauge group etc...

There are two remarks we would like to make concerning these types
of mirror fermions. First, experimental constraints on mirror
quarks and leptons masses are not very well studied since
they depend on the specifics of decay modes. However, as
discussed in \cite{hung1}, one can safely conclude that the
right-handed neutrinos have to be more massive than $M_Z/2$ because
of constraints from the width of the Z boson. The lower bound
on the charged mirror lepton masses can be taken to be around
$\sim 100$ GeV (long-lived heavy leptons) \cite{PDG} if not lower.
The bounds on the mirror quarks are murkier but one can assume
that they might be heavier than $\sim 200$ GeV (from bounds on
long-lived quarks) \cite{PDG}. The second remark has to do with
electroweak precision parameter constraints. As mentioned in \cite{hung1}, 
the positive contribution of mirror fermions
to the S parameter can be compensated by a negative contribution from
the Higgs triplet sector present in the model \cite{hung3}. 
In \cite{hung3}, it was shown that, depending on the mass splitting
inside the triplet Higgs scalar(s), its contribution to S can be negative
and relatively large in magnitude, and can offset the positive
contributions coming from the mirror fermions. Furthermore,
it is well-known that Majorana fermions can also have
a negative contribution to S \cite{dugan}.
Notice that
it is straightforward to satisfy the $\rho$-parameter constraint
\cite{hung1}. The above remarks and other phenomenological issues
concerning mirror fermions will be dealt with elsewhere.

Since $\nu_R$'s are now members of $SU(2)$ doublets, it was shown in
\cite{hung1} that the Majorana masses $M_R$  coming from the 
term $M_R \, \nu_{R}^{T}\, \sigma_2 \, \nu_{R}$
is now related to the electroweak breaking scale $\Lambda_{EW} \sim 246\,GeV$, 
rendering the detection of a low-scale $\nu_R$ at colliders and observing
lepton-number violating processes a real possibility. In this scenario,
one can directly test the see-saw mechanism at collider energies as
well as the Majorana nature of $\nu_R$'s. 

The Dirac mass term now
comes from the {\em mixing} between the SM left-handed doublet
and the mirror right-handed doublet through the coupling with
a singlet scalar field which develops a non-vanishing vacuum-expectation value
(VEV) $v_S$, giving $m_D = g_{Sl}\,v_S$, where $g_{Sl}$ is
the Yukawa coupling. The light neutrinos becomes massless in the
limit $g_{Sl} \rightarrow 0$, i.e. when the SM particles and their
mirror counterparts decouple. It is this mixing which
will prove important for the LFV processes considered in this
paper. 

What is parity violation? Basically, this is equivalent to
the question of why the weak interactions of SM particles
are of the V-A type. This rather old question never
really disappears and occasionally has new twists to it,
especially in light of the upcoming explorations by
the LHC and hopefully also by the ILC.
For the SM, this is built in because of
the absence of right-handed neutrinos.
In the LR model, this is due to $M_{W_R} > M_{W_L}$.
Parity is restored for $E \gg M_{W_R}$.
For $E \gg M_{W_R} > M_{W_L}$, the V+A interactions have equal
strength to that of the V-A interactions and parity is restored. 

In the model of electroweak-scale right-handed neutrinos,
the question of parity violation takes on a slightly different
meaning as compared with the LR model. Here,  
$SU(2) \otimes U(1)$ as a gauge theory is actually a vector-like model
in the sense that fermions of both chiralities are present.
In consequence, ``parity restoration'' refers to the
existence of the mirror fermions in our model. 
Furthermore, the SM fermions have V-A weak interactions at tree-level
but, because of the existence of a {\em mixing} between SM and mirror
fermions, receive V+A interactions through one-loop diagrams
which are of course suppressed with respect to the tree-level
V-A interactions. 
As with the LR model, these (radiative) V+A
contributions vanishes in the limit $g_{Sl} \rightarrow 0$
which also implies a vanishing Dirac mass $m_D$ and hence
a vanishing light neutrino mass.
Here, one expects deviations of
the SM couplings of the SM quarks and leptons due to the
mixing between SM fermions and their mirror counterparts.
This includes corrections to the electroweak processes
involving electrons and neutrinos as well as lepton flavor
violating processes such as $\mu \rightarrow e\, \gamma$
and $\tau \rightarrow \mu\,\gamma$.
It is the latter processes that we will concentrate on in this
paper. Details of the former processes will be presented elsewhere
\cite{hung4}.

We will see below that there is a deep connection
between the mixing parameters involved in LFV processes and
the Dirac neutrino mass matrix which participates in the seesaw
mechanism. Since these same parameters \cite{hung1} participate in the
determination of the decay length of the mirror charged leptons
, $l^{M}_R$, it will be shown below that were these LFV
processes to be detected the decay length would be microscopic.
If the charged mirror leptons were to be found with ``macroscopic''
decay lengths, LFV processes would be practically unobservable in 
our model. One word of reminder is in order here on why
it is important to gain a good understanding on the
size of the decay lengths of the mirror leptons. Because
the model presented in \cite{hung1} deals with right-handed
neutrinos with electroweak-scale masses, it is possible
to probe the Majorana nature of $\nu_R$'s and the see-saw
mechanism at present and future colliders (Tevatron, LHC, ILC).
The telltale signatures would be SM like-sign dileptons \cite{keung} which
are produced e.g. in processes such as
 \begin{equation}
\label{production}
q + \bar{q} \rightarrow Z
\rightarrow  \nu_R + \nu_R \,.
\end{equation}
Since $\nu_R$'s are Majorana particles, they can have transitions
such as $\nu_R \rightarrow l^{M,\mp}_R + W^{\pm}$.
A heavier $\nu_R$ can decay into a lighter $l^{M}_R$ and one can have
\ba
\label{production2}
\nu_R + \nu_R &\rightarrow& l^{M,\mp}_R +l^{M,\mp}_R +W^{\pm} + W^{\pm} \nonumber \\
&\rightarrow& l^{\mp}_L +l^{\mp}_L +W^{\pm} + W^{\pm}
+ \phi_S + \phi_S \,,
\ea
where $l^{M,\mp}_R \rightarrow l^{\mp}_L + \phi_S$ and
where $\phi_S$, the singlet scalar field in the model, would
constitute the missing energy.

\section{Brief review of the electroweak scale right-handed
neutrino model}
\label{model}

In our model, the dominant contribution to the process
$l_{i} \rightarrow l_{j}\, \gamma$ comes from the diagrams shown
in Fig. 1. The Feynman rules for the diagrams can be extracted from
the Lagrangians given below.
Let us first write down the Lagrangian of our model and
enumerate the particle content and various symmetries.

The gauge group is $SU(2) \otimes U(1)$.

Below is a list of the fermion content. 
Because of the problem at hand, namely the processes
$\mu \rightarrow e\, \gamma$ and
$\tau \rightarrow \mu\,\gamma$, we shall concentrate
mainly on the leptons in this paper.


\bi

\item $(2, Y/2=-1/2)$ (Lepton doublets):

\begin{equation}
\label{sm}
l_L = \left( \begin{array}{c}
\nu_L \\
e_{L}
\end{array} \right)_{i} \,,
\end{equation}

\begin{equation}
\label{mirror}
l^{M}_R = \left( \begin{array}{c}
\nu^{M}_R \\
e^{M}_{R}
\end{array} \right)_{i} \,,
\end{equation}
where $i$ stands for the family number. In \cite{hung1}, $\nu^{M}_R$ is identified with
the right-handed neutrino, a fact which gives rise to an electroweak-scale right-handed
neutrino.


\item $(1, Y/2=-1)$ (Lepton singlets):

\be
\label{lsinglet}
e_{iR} \,;\, e^{M}_{iL} \,.
\ee

\ei

Next, we write down the Lagrangian involving the mirror leptons and their interactions
with the SM leptons. In so doing, we will
also include for clarity the term that gives the Majorana mass to the
right-handed neutrinos although it is not needed for the present discussion.
Also for completeness we will write down the Lagrangian for the SM leptons in
order to make a comparison.
We have the following interactions.
\bi

\item Charged current interactions:
\be
\label{lagrangian1}
{\cal L}^{CC}= {\cal L}^{CC}_{SM} + {\cal L}^{CC}_{M} \,,
\ee
where
\ba
\label{lagrangian2}
{\cal L}^{CC}_{SM}&=& -(\frac{g}{2\,\sqrt{2}})\sum_{i}\,\bar{\psi}^{SM}_{i}\,\gamma^{\mu}\,
(1-\gamma_{5})[T^{+}\,W^{+}_{\mu} \nonumber \\ 
&& + T^{-}\,W^{-}_{\mu}]\,\psi^{SM}_{i} \,,
\ea
\ba
\label{lagrangian3}
{\cal L}^{CC}_{M}&=& -(\frac{g}{2\,\sqrt{2}})\sum_{i}\,\bar{\psi}^{M}_{i}\,\gamma^{\mu}\,
(1+\gamma_{5})[T^{+}\,W^{+}_{\mu}  \nonumber \\
&&+ T^{-}\,W^{-}_{\mu}]\,\psi^{M}_{i} \,,
\ea
and
\begin{equation}
\label{psism}
\psi^{SM} = \left( \begin{array}{c}
\nu \\
e
\end{array} \right)_{i} \,,
\end{equation}
\begin{equation}
\label{psim}
\psi^{M} = \left( \begin{array}{c}
\nu^{M} \\
e^{M}
\end{array} \right)_{i} \,.
\end{equation}
Above the subscripts $SM$ and $M$ refer to SM and mirror particles respectively.
Notice that Eq. (\ref{lagrangian2}) has $(1-\gamma_{5})$ as opposed to
$(1+\gamma_{5})$ of Eq. (\ref{lagrangian3}).

\item Neutral current interactions:
\be
\label{lagrangian4}
{\cal L}^{NC}= {\cal L}^{NC}_{SM} + {\cal L}^{NC}_{M} \,,
\ee
where
\ba
\label{lagrangian5}
{\cal L}^{NC}_{SM}&=& -(\frac{g}{4\,\cos \theta_{W}})\,Z_{\mu}\,\{\sum_{i} \bar{\nu}_{i}\,
\gamma^{\mu}\,(1-\gamma_{5})\,\nu_{i}  \nonumber \\
&&+ \sum_{i}
\bar{e}_{i} \gamma^{\mu}[(-1+4\,\sin^{2}\theta_{W})+\gamma_5]e_{i}\} \,,
\ea
\ba
\label{lagrangian6}
{\cal L}^{NC}_{M}&=& -(\frac{g}{4\,\cos \theta_{W}})\,Z_{\mu}\,\{\sum_{i} \bar{\nu}^{M}_{i}\,
\gamma^{\mu}\,(1+\gamma_{5})\,\nu^{M}_{i}  \nonumber \\
&&+ \sum_{i}
\bar{e}^{M}_{i} \gamma^{\mu}[(-1+4\,\sin^{2}\theta_{W})-\gamma_5]e^{M}_{i}\}.
\ea

\item Electromagnetic interactions:
\be
\label{em}
{\cal L}^{EM}= e\,\sum_{i}(\bar{e}_{i} \,\gamma^{\mu}\,e_{i} + \bar{e}^{M}_{i} \,\gamma^{\mu}\,
e^{M}_{i})\,A_{\mu} \,.
\ee

\item Yukawa interactions which contribute to the neutrino Dirac mass term:

This is the most important part which contributes to the aforementioned
LFV rare processes. The gauge-invariant Yukawa Lagrangian can be
written as follows:

\ba
\label{yuk1}
{\cal L}_S &=& - \bar{l}^{0}_{L}\,g_{Sl}  \, l^{0,M}_{R}\,\phi_S + H.c. 
\nonumber \\
&=& - (\bar{\nu}^{0}_{L}\, g_{Sl}\, \nu^{0,M}_{R} + 
\bar{e}^{0}_{L} \,g_{Sl}\, e^{0,M}_{R})\,\phi_S
+ H.c \,,
\ea
where $\phi_S$ is the singlet scalar field whose vacuum expectation value (VEV)
gives rise to the neutrino Dirac mass term \cite{hung1}. This is combined
with the Majorana mass term for the right-handed neutrinos in a see-saw mechanism
as shown in \cite{hung1}. In (\ref{yuk1}), $\nu^{0}_L$, $\nu^{0,M}_R$, 
$e^{0}_L$, and $e^{0,M}_R$ denote column vectors with $n$ components for
$n$ families and $g_{Sl}$ denotes an $n \times n$ coupling matrix. From
hereon, we will take for definiteness $n=3$.

\item Yukawa interactions involving $SU(2)_L$ singlets, $e_R$ and $e^{M}_L$:
\be
\label{yuk2}
{\cal L}^{'}_S= - (\bar{e}^{0}_{R} \,g^{'}_{Sl}\, e^{0,M}_{L})\,\phi_S + H.c. \,,
\ee 
where $g^{'}_{Sl}$ is a $3 \times 3$ matrix for 3 families.

\item Yukawa interactions giving rise to the Majorana mass for the right-handed
neutrinos:
\be
\label{majorana}
{\cal L}_M = l^{M,T}_{R}\, \sigma_2 \,\tau_2 \,g_{M}\,\tilde{\chi}\, l^{M}_{R}\,,
\ee
where 
\be
\label{delta}
\tilde{\chi} = \frac{1}{\sqrt{2}}\,\vec{\tau}.\vec{\chi}=
\left( \begin{array}{cc}
\frac{1}{\sqrt{2}}\,\chi^{+} & \chi^{++} \\
\chi^{0} & -\frac{1}{\sqrt{2}}\,\chi^{+}
\end{array} \right) \,,
\ee \\
and $g_{M}$ is a $3 \times 3$ matrix for 3 families.
Notice that Eq. (\ref{majorana}) is written down for completeness in this manuscript. As shown
in \cite{hung1}, this term gives rise to an electroweak-scale Majorana mass for the right-handed
neutrinos when the $SU(2)_L$-triplet Higgs field develops a VEV, $\langle \chi^{0}\rangle = v_M$, 
of the order $O(\Lambda_{EW})$. 
(The subtleties associated with such a large triplet VEV are discussed in \cite{hung1}.)
As of now, there is no direct evidence for Higgs triplets, and for that matter, also
for Higgs doublet(s). However, one of the characteristics of this model is
the existence of particles such as doubly-charged Higgs bosons. Present limits
of around $100$ GeV are rather model-dependent. A study of the Higgs sector of our
model is under investigation \cite{hung5}.

\ei

The above mixings between the SM fermions and their
mirror counterparts also give rise to mass mixings in the charged lepton
sector (as well as in the quark sector). A brief review of the points made in
\cite{hung1} goes as follows.
Take, for example, one family
of fermions. When $\phi_S$ develops a VEV $v_S$, one obtains a Dirac mass
for the neutrino and a mass mixing between the SM and mirror charged
leptons as follows (assuming $g^{'}_{Sl} = g_{Sl}$)
\be
\label{neumass}
m_{\nu}^D = g_{Sl} \, v_S \,,
\ee
for the Dirac neutrino and
\be
\label{charged}
M_l = \left( \begin{array}{cc}
m_l& m_{\nu}^D \\
m_{\nu}^D & m_{l^M}
\end{array} \right) \,,
\ee
for the charged SM and mirror leptons. In Eq. (\ref{charged}), $m_l$ is
a generic notation for the mass of a SM charged lepton obtained
from the coupling to the SM Higgs doublet. Similarly,
$m_{l^M}$ is a generic notation for the mirror charged lepton mass
obtained in the same manner. The off-diagonal element in Eq. (\ref{charged})
comes from the cross coupling to the singlet scalar field through
Eq. (\ref{yuk1}).
The diagonalization of (\ref{charged}) gives the following eigenvalues
for the charged (SM and mirror) lepton masses
\bes
\be
\label{ml}
\tilde{m}_l = m_l\, - \, \frac{(m_{\nu}^{D})^2}{m_{l^M} - m_l}  
\ee
\be
\label{mlm}
\tilde{m}_{l^M} = m_{l^M} \, + \, \frac{(m_{\nu}^{D})^2}{m_{l^M} - m_l} \,.
\ee
\ees
As shown in \cite{hung1}, the Majorana mass of the right-handed neutrino
$M_R = g_M\,v_M$, where
$v_M \sim O(\Lambda_{EW})$. The seesaw mechanism then
gives a light neutrino a mass $(m_{\nu}^{D})^2/M_R$ of $O(< 1\,eV)$
and a heavy neutrino of mass $M_Z /2 <M_R < O(\Lambda_{EW})$. 
This makes the right-handed neutrinos detectable by future collider
experiments.
This also implies
that $m_{\nu}^{D} \sim 10^{5}\,eV$. Since the mirror lepton mass is
of the order of the electroweak scale \cite{hung1}, it follows that the
second terms of the right-hand side of (\ref{ml}, \ref{mlm}) are
tiny compared with the first terms and can be ignored.
Hence, for all practical purposes,
the masses of the charged fermions are those obtained in a SM way i.e.
through the Yukawa couplings with the SM Higgs doublet(s).
As a result, the mass eigenstates
for the charged (SM and mirror) leptons are principally obtained
in this way. This will be the (rather good) approximation that we will
use below in the discussion of LFV rare processes involving charged leptons.

In what follows, we will make use of the remarks made above
concerning mass eigenstates for the charged leptons. 
To express (\ref{yuk1}) in terms of mass eigenstates, let us define
\be
\label{eigen1}
e^{0}_L = U^{l}_L e_L \,;\, e^{0,M}_R = U^{l^M}_R e^{M}_R \,.
\ee
Using (\ref{eigen1}), one can now rewrite the charged lepton part of (\ref{yuk1})
in terms of the mass eigenstates as follows
\be
\label{yul1p}
{\cal L}_{S,charged} =  - (\bar{e}_{L} \,U^{L}\,e^{M}_{R})\,\phi_S + H.c. \,,
\ee
 where
\be
\label{UlL}
U^{L} = U^{l,\dag}_L \, g_{Sl}\, U^{l^M}_R \,.
 \ee

From \cite{hung1}, one can deduce from (\ref{yuk1}) the Dirac mass matrix of the 
neutrino sector when $\langle \phi_S \rangle = v_S$ as follows
\be
\label{numatrix}
m^{D}_{\nu} = v_S \, g_{Sl} \,.
\ee
In terms of (\ref{numatrix}), the matrix $U^{L}$ can now be written as
\be
\label{UlL2}
U^{L} = U^{l,\dag}_L \,(\frac{m^{D}_{\nu}}{v_S})\, U^{l^M}_R \,.
 \ee
One can see that the matrix $U^{L}$ which mixes different families
of SM charged leptons with those of the mirror leptons now involves
the Dirac neutrino mass matrix. LFV processes to be
discussed in this paper will in consequence indirectly probe the Dirac
part of the neutrino mass matrix. In fact, one can invert Eq. (\ref{UlL2})
to obtain
\be
\label{mD1}
\frac{m^{D}_{\nu}}{v_S} = U^{l}_L\,U^{L}\,U^{l^M, \dag}_R \,.
\ee
As discussed in \cite{hung1}, let
us recall that the see-saw mechanism yields the following
mass matrix for the light neutrino sector
\be
\label{lightnu}
m_{\nu,light}= - m^{D,T}_{\nu}\, M^{-1}_{R}\, m^{D}_{\nu}\,,
\ee 
while the mass matrix for the heavy right-handed sector
is simply $M_R$. Once again, let us notice that
$U^{L} \rightarrow 0$ in the limit 
$m_{\nu,light} (m^{D}_{\nu}) \rightarrow 0$, and
there will be no mixing between SM and mirror fermions.

For the $SU(2)_L$ singlets, one has
\be
\label{eigen3}
e^{0}_R = U^{l}_R e_R \,;\, e^{0,M}_L = U^{l^M}_L e^{M}_L \,,
\ee
(\ref{yuk2}) is rewritten as
\be
\label{yuk2p}
{\cal L}^{'}_S= - (\bar{e}_{R} \,U^{R} \, e^{M}_{L})\,\phi_S  + H.c. \,,
\ee 
where
\be
\label{UlR}
U^{R} = U^{l,\dag}_R \,g^{'}_{Sl}\, U^{l^M}_L \,.
\ee
Although it is not necessary to do so, one can further simplify the problem
by assuming that $g^{'}_{Sl} = g_{Sl}$ in which case we obtain
\be
\label{UlR2}
U^{R} = U^{l,\dag}_R \,(\frac{m^{D}_{\nu}}{v_S})\, U^{l^M}_L \,.
\ee
Similarly to (\ref{mD1}), one can invert (\ref{UlR2}) to obtain
\be
\label{mD2}
\frac{m^{D}_{\nu}}{v_S} = U^{l}_R\,U^{R}\,U^{l^M, \dag}_L \,.
\ee

From the above discussion, one cannot fail but notice the deep connection
between the Dirac part of the neutrino mass matrix $m^{D}_{\nu}$ and the
matrices which are involved in LFV processes in our model, namely $U^{L}$
and $U^{R}$.

\section{The processes $\mu \rightarrow e\, \gamma$ and $\tau \rightarrow \mu\,\gamma$}

The processes $\mu \rightarrow e\, \gamma$ and $\tau \rightarrow \mu\,\gamma$ 
can now be computed in our model by
using the interaction Lagrangians listed above. The Feynman rules needed for
Fig. 1 can be read
off Eqs. (\ref{lagrangian3},\ref{lagrangian6}, \ref{em}, \ref{yul1p}, \ref{yuk2p}).
Notice that there is also a contribution to these LFV processes coming
from diagrams with a W and a light neutrino propagating inside
the loop. But this is entirely negligible as noticed by \cite{chengli}.

Fig. 1 is the diagram which contains a magnetic moment term (i.e.
proportional to $\sigma_{\mu\,\nu}$) and will
be the one that we concentrate on. (Two other diagrams with
the photon line attached to the external legs need not be considered,
being proportional to $\gamma_{\mu}$, see e.g.\cite{muegam}.)
\begin{figure}
\includegraphics[angle=0,width=6cm]{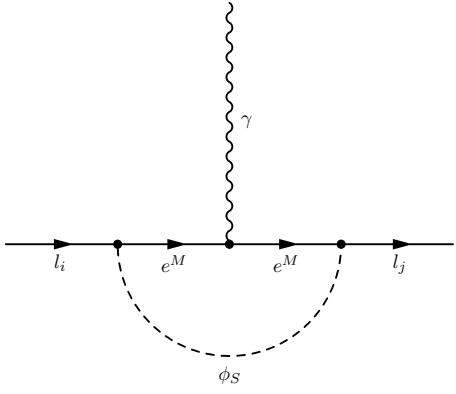}
\caption{Dominant diagram for the LFV process $l_i \rightarrow l_j + \gamma$}
\end{figure}

The general form of the amplitude can be written as
\ba
\label{amplitude}
T(l_i \rightarrow l_j\, \gamma) &=& \epsilon^{\lambda} \bar{u}_{l_j}(p-q)
\{i\,q^{\nu} \, \sigma_{\lambda \nu} \,[c^{(l_i)}_{L}(\frac{1-\gamma_5}{2}) \nonumber \\
&& +c^{(l_i)}_{R}(\frac{1+\gamma_5}{2})]\}u_{l_i}(p) \,.
\ea
The decay rate  and the branching ratio are respectively
\be
\label{rate}
\Gamma(l_i \rightarrow l_j\, \gamma) = \frac{m_{l_i}^3}{16\,\pi}(|c^{(l_i)}_{L} + c^{(l_i)}_{R}|^2 + 
|c^{(l_i)}_{L} - c^{(l_i)}_{R}|^2) \,,
\ee
\ba
\label{branchingmu}
B(\mu \rightarrow e\, \gamma) &=& \frac{\Gamma(\mu \rightarrow e\, \gamma)}
{\Gamma(\mu \rightarrow e \bar{\nu}_e \nu_{\mu})} \nonumber \\
&=&\frac{12\pi^2}{m_{\mu}^2 G_{F}^2}(|c^{(\mu)}_{L} + c^{(\mu)}_{R}|^2 + 
|c^{(\mu)}_{L} - c^{(\mu)}_{R}|^2) \,. \nonumber \\
\ea
\ba
\label{branchingtau}
\frac{B(\tau \rightarrow \mu\, \gamma)}{B(\tau \rightarrow \mu \bar{\nu}_{\mu} \nu_{\tau})} 
&=& \frac{\Gamma(\tau \rightarrow \mu\, \gamma)}
{\Gamma(\tau \rightarrow \mu \bar{\nu}_{\mu} \nu_{\tau})} \nonumber \\
&=&\frac{12\pi^2}{m_{\tau}^2 G_{F}^2}(|c^{(\tau)}_{L} + c^{(\tau)}_{R}|^2 + 
|c^{(\tau)}_{L} - c^{(\tau)}_{R}|^2) \,. \nonumber \\
\ea
$c^{(\mu)}_{L}$, $c^{(\mu)}_{R}$, $c^{(\tau)}_{L}$ and $c^{(\tau)}_{R}$ are computed from the diagram shown in Fig. 1
to be
\be
\label{cl}
c^{(\mu)}_{L} = \frac{1}{64\,\pi^2} \sum_{i} \frac{U^{R*}_{i\mu}\,U^{L}_{ei}}{m_i}\,,
\ee
\be
\label{cr}
c^{(\mu)}_{R} = \frac{1}{64\,\pi^2} \sum_{i} \frac{U^{L*}_{i\mu}\,U^{R}_{ei}}{m_i}\,,
\ee
\be
\label{cl2}
c^{(\tau)}_{L} = \frac{1}{64\,\pi^2} \sum_{i} \frac{U^{R*}_{i\tau}\,U^{L}_{\mu i}}{m_i}\,,
\ee
\be
\label{cr2}
c^{(\tau)}_{R} = \frac{1}{64\,\pi^2} \sum_{i} \frac{U^{L*}_{i\tau}\,U^{R}_{\mu i}}{m_i}\,,
\ee
where $m_i$ are the masses of the charged mirror leptons. (In obtaining the above results,
we looked at the coefficient of the term $p.\epsilon$, where $p$ is the momentum of the
decaying particle and $\epsilon$ is the polarization of the photon. Also, as discussed
in \cite{hung1}, $m_S \ll m_i$.)
Notice in (\ref{branchingmu}) and 
(\ref{branchingtau}) that whereas $B(\mu \rightarrow e \bar{\nu}_e \nu_{\mu}) \sim 100\%$,
one has $B(\tau \rightarrow \mu \bar{\nu}_{\mu} \nu_{\tau}) \sim 17.4 \%$.

A few special cases are worth noticing.

\bi

\item $g_{Sl} = g^{'}_{Sl}$, $U^{l}_{L} = U^{l}_{R}$ and $U^{l^M}_{R} = U^{l^M}_{L}$:

The last two above equalities simply imply that one {\em assumes} that the mass matrices
for the SM charged leptons and those of the mirror leptons are such that the
``left'' and ``right'' diagonalization matrices are {\em identical}. With these assumptions,
one obtains
\be
\label{ULUR}
U^{L} = U^{R} = U^{E} \,. 
\ee
With these assumptions, one obtains
\bes
\label{clp1}
\be
c^{(\mu)}_{L} = c^{(\mu)}_{R}= \frac{1}{64\,\pi^2} \sum_{i} \frac{U^{E*}_{i\mu}\,U^{E}_{ei}}{m_i}\,,
\ee
\be
c^{(\tau)}_{L} = c^{(\tau)}_{R}= \frac{1}{64\,\pi^2} \sum_{i} \frac{U^{E*}_{i\tau}\,U^{E}_{\mu i}}{m_i}\,.
\ee
\ees
It is also convenient to rewrite the above expressions in terms of a mass $m_E$ (more like an
average of the three mirror lepton masses) as follows
\bes
\label{clp2}
\be
c^{(\mu)}_{L} = c^{(\mu)}_{R}= \frac{1}{64\,\pi^2} \frac{1}{m_E}\sum_{i} (\frac{m_E}{m_i})(U^{E*}_{i\mu}\,U^{E}_{ei})\,,
\ee
\be
c^{(\tau)}_{L} = c^{(\tau)}_{R}= \frac{1}{64\,\pi^2} \frac{1}{m_E}\sum_{i} (\frac{m_E}{m_i})(U^{E*}_{i\tau}\,U^{E}_{\mu i})\,.
\ee
\ees

What are the implications of the above assumption?

\item Degenerate charged mirror leptons: $m_i = m_E$.

In this case
one is left with the factors $\sum_{i} U^{E*}_{i\mu}\,U^{E}_{ei}$ and
$\sum_{i} U^{E*}_{i\tau}\,U^{E}_{\mu i}$ which would {\em vanish
identically} if $g_{Sl}$ (or equivalently $m^{D}_{\nu}/v_S$) were proportional to
the unit matrix because in this case the matrix $U^{E}$ would be {\em unitary}.
One would then have $c^{(\mu)}_{L} = c^{(\mu)}_{R}=0$
and $c^{(\tau)}_{L} = c^{(\tau)}_{R}=0$, implying
$B(\mu \rightarrow e\, \gamma)=0$ and $B(\tau \rightarrow \mu\, \gamma)=0$. 
However $U^{E}$ would no longer be {\em unitary} if $m^{D}_{\nu}/v_S$
were {\em not} proportional to the unit matrix, a fact which implies a non-vanishing
value for the aforementioned branching ratios.
Once again, one notices the implication of the form of neutrino mass matrices on LFV
processes.


\item Non-degenerate charged mirror leptons: $m_1=m_E$, $m_2=m_E+\delta m_2$, and 
$m_3=m_E+\delta m_3$ with the assumption $|\delta m_{2,3}| \ll m_E$.

The non-degenerate case implies that the above branching ratios can be non-vanishing
even if $m^{D}_{\nu}/v_S$ were proportional to the unit matrix. 




\ei

\section{Constraints on the model from $B(\mu \rightarrow e\, \gamma)_{exp}$ and 
$B(\tau \rightarrow \mu\, \gamma)_{exp}$}

The experimental constraints we will use here are those from BaBar \cite{bbar} 
and Belle \cite{belle}
for $B(\tau \rightarrow \mu\, \gamma)_{exp}$ and from PDG \cite{PDG} for
$B(\mu \rightarrow e\, \gamma)_{exp}$. They are
$B(\tau \rightarrow \mu\, \gamma)_{exp} < 6.8 \times 10^{-8}$ for BaBar \cite{bbar} 
and $B(\tau \rightarrow \mu\, \gamma)_{exp} < 4.5 \times 10^{-8}$ for Belle \cite{belle};
$B(\mu \rightarrow e\, \gamma)_{exp} < 1.2 \times 10^{-11}$ \cite{PDG}. Also
we will be using $B(\tau \rightarrow \mu \bar{\nu}_{\mu} \nu_{\tau}) \sim 17.4 \%$.

Although the most general discussion would involve the non-degenerate
case with $U^{L} \neq U^{R}$, it is probably more illuminating to
first investigate the non-degenerate scenario with $U^{L} = U^{R}$. For
this, we will use Eq. (\ref{clp2}). We obtain for
$B(\mu \rightarrow e\, \gamma)$
\ba
\label{branchingmu2}
B(\mu \rightarrow e\, \gamma) &=& (\frac{3}{256\,\pi^2})
(\frac{1}{m_{\mu}^2 G_{F}^2 m_E^2}) \nonumber \\
&& \times|\sum_{i} (\frac{m_E}{m_i})(U^{E*}_{i\mu}\,U^{E}_{ei})|^2 . 
\nonumber \\
\ea
and for $B(\tau \rightarrow \mu\, \gamma)$
\ba
\label{branchingtau2}
B(\tau \rightarrow \mu\, \gamma) &=&(\frac{3}{256\,\pi^2})(\frac{1}{m_{\tau}^2 G_{F}^2 m_E^2}) \times 0.174 \nonumber \\
&& \times |\sum_{i} (\frac{m_E}{m_i})(U^{E*}_{i\tau}\,U^{E}_{\mu i})|^{2} .
\nonumber \\
\ea
It is also useful to relate one branching ratio to another as follows
\ba
\label{branchratio}
B(\mu \rightarrow e\, \gamma) &=&
\frac{|\sum_{i} (\frac{m_E}{m_i})(U^{E*}_{i\mu}\,U^{E}_{ei})|^2}
{|\sum_{i} (\frac{m_E}{m_i})(U^{E*}_{i\tau}\,U^{E}_{\mu i})|^{2}} \nonumber \\
&& \times \frac{B(\tau \rightarrow \mu\, \gamma)}{0.174} \times (\frac{m_{\tau}}{m_{\mu}})^2\,.
\ea

We will illustrate the previous results with two examples: $m_E = 100$ GeV and $m_E = 200$ GeV. 
\bi

\item $m_E = 100$ GeV:

\be
\label{boundmu}
|\sum_{i} (\frac{m_E}{m_i})(U^{E*}_{i\mu}\,U^{E}_{ei})|^2
< 1.25 \times 10^{-15} \,,
\ee
\be
\label{boundtau}
|\sum_{i} (\frac{m_E}{m_i})(U^{E*}_{i\tau}\,U^{E}_{\mu i})|^{2}
<\left\{ \begin{array}{c}
7.1 \\ 
4.7 \end{array}
\right\} \times 10^{-12}  \,,
\ee
where the first and second numbers on the right-hand-side of Eq. (\ref{boundtau}) refer to BaBar and Belle respectively. 

\item $m_E = 200$ GeV:

\be
\label{boundmu2}
|\sum_{i} (\frac{m_E}{m_i})(U^{E*}_{i\mu}\,U^{E}_{ei})|^2
< 5.0 \times 10^{-15} \,,
\ee
\be
\label{boundtau2}
|\sum_{i} (\frac{m_E}{m_i})(U^{E*}_{i\tau}\,U^{E}_{\mu i})|^{2}
<\left\{ \begin{array}{c}
28.4 \\ 
18.8 \end{array}
\right\} \times 10^{-12}  \,.
\ee

\ei

As we have mentioned in the beginning of the manuscript, the matrix elements of $U^{E}$ determine
the lifetime and hence the decay length of the mirror charged leptons. Also, as we have seen
above, some of these elements are constrained by the LFV processes 
$\mu \rightarrow e\, \gamma$ and $\tau \rightarrow \mu\, \gamma$. Let us make some rough assumptions on the 
above bounds in order to gain some insights into what one might expect.

Let us take the case $m_E =100\,GeV$ for definiteness and let us assume:
\be
\label{assumption1}
 |\delta m_{i}| \ll m_E\,.
\ee
Let us also assume the following hierarchy (Case I) with $i=1,2,3$:
\be
\label{assumption2}
U^{E}_{i\,e} \sim \lambda^3\,;\,U^{E}_{i\,\mu} \sim \lambda^2\,;\,U^{E}_{i\,\tau} \sim \lambda\,.
\ee
With the above assumptions, one can now rewrite the bounds (\ref{boundtau}, \ref{boundmu})
(taking e.g. the Belle value) as
\be
\label{newbound1}
\lambda^6 <  5.2 \times 10^{-13} \,, 
\ee
\be
\label{newbound2}
\lambda^{10}<  1.4 \times 10^{-16} \,
\ee
where (\ref{newbound1}) refers to the bound from $\tau \rightarrow \mu\, \gamma$
while (\ref{newbound2}) comes from the bound on $\mu \rightarrow e\, \gamma$.
(\ref{newbound1}) gives $\lambda <0.009$.
This gives $\lambda^{10} < 10^{-21}$ which satisfies (\ref{newbound2}).
Suppose that the process $\tau \rightarrow \mu\, \gamma$
can be experimentally probed with a branching
ratio not too far below the current bound. What does it say about the
detectability of $\mu \rightarrow e\, \gamma$?
This can be easily estimated by looking at the relation (\ref{branchratio})
\be
\label{branchratio2}
B(\mu \rightarrow e\, \gamma) \approx 1.6 \times 10^{3} \, \lambda^4 \,B(\tau \rightarrow \mu\, \gamma) \,.
\ee
With $\lambda <0.009$, one would have 
$B(\mu \rightarrow e\, \gamma) \sim 10^{-8} B(\tau \rightarrow \mu\, \gamma) < 4 \times 10^{-16}$ which makes
the process $\mu \rightarrow e\, \gamma$ unobservable in the near future, about two orders
of magnitude below the sensitivity of the MEG proposal \cite{MEG}.

A more interesting hierarchy (Case II) is as follows: ($i=1,2,3$)
\be
\label{assumption3}
U^{E}_{i\,e} \sim \lambda^3\,;\,U^{E}_{i\,\mu} \sim \lambda\,;\,U^{E}_{i\,\tau} \sim \lambda^2\,.
\ee
We now obtain the following bounds for (\ref{boundtau}, \ref{boundmu})
\be
\label{newbound3}
\lambda^6 <  5.2 \times 10^{-13} \,, 
\ee
\be
\label{newbound4}
\lambda^{8}<  1.4 \times 10^{-16} \,.
\ee

Note that (\ref{newbound3}) is identical to (\ref{newbound1}) since one is simply switching
the role of $\mu$ and $\tau$, but now the exponent on the left-hand-side of (\ref{newbound4})
is lower $8$ instead of $10$. Again (\ref{newbound3}) gives $\lambda <0.009$. One now
has $\lambda^{8} < 4.3 \times 10^{-17}$, a factor of three below
(\ref{newbound4}) which is clearly satisfied. There is
in addition a {\em huge} advantage: If $\tau \rightarrow \mu\, \gamma$ is {\em discovered}
e.g. slightly below the present bound then one {\em might expect}
to discover $\mu \rightarrow e\, \gamma$ with a rate of about a factor of three below
its present experimental limit! In fact,the relationship (\ref{branchratio}) is now
\be
\label{branchratio3}
B(\mu \rightarrow e\, \gamma) \approx 1.6 \times 10^{3} \, \lambda^2 \,B(\tau \rightarrow \mu\, \gamma) \,.
\ee
With $\lambda <0.009$, one now has
$B(\mu \rightarrow e\, \gamma) \sim 10^{-4} B(\tau \rightarrow \mu\, \gamma) < 5 \times 10^{-12}$.
The observability of one process implies that of the
other. Note that the estimated bound on $B(\mu \rightarrow e\, \gamma)$ is well within
the range of the MEG proposal \cite{MEG}.

Last but not least, we could also consider the ``inverted hierarchy'' scenario (Case III):
\be
\label{assumption4}
U^{E}_{i\,e} \sim \lambda\,;\,U^{E}_{i\,\mu} \sim \lambda^2\,;\,U^{E}_{i\,\tau} \sim \lambda^3\,.
\ee
This case results in for (\ref{boundtau}, \ref{boundmu})
\be
\label{newbound5}
\lambda^{10} <  5.2 \times 10^{-13} \,, 
\ee
\be
\label{newbound6}
\lambda^{6}<  1.4 \times 10^{-16} \,.
\ee
This case is the mirror of Case I. Using (\ref{newbound5}), one obtains
$\lambda <0.06$. For $\lambda$ close to this upper limit, (\ref{newbound6}) cannot be
satisfied. If we satisfy (\ref{newbound6}) with $\lambda <0.002$,
$B(\tau \rightarrow \mu\, \gamma)$ which is proportional to $\lambda^{10} <10^{-27}$
will be hopelessly small. 

The above constraints (Cases I, II, and III) are listed for
convenience in Table I. Notice that the upper bounds on $\lambda$
listed in Table I are those which satisfy the experimental constraints
from {\em both} $B(\tau \rightarrow \mu\, \gamma)$ and $B(\mu \rightarrow e\, \gamma)$.
\begin{table}
\caption{Bounds on branching ratios for three mixing scenarios (\ref{assumption2}),
(\ref{assumption3}), (\ref{assumption4})} 
\begin{ruledtabular}
\begin{tabular}{cccc} 
&$\lambda \leq$&$B(\tau \rightarrow \mu\, \gamma) \leq$ &$B(\mu \rightarrow e\, \gamma) \leq$ \\ \hline
Case I: (\ref{assumption2})&0.009 & 
$4.5 \times 10^{-8}$& $4 \times 10^{-16}$\\ \hline
Case II: (\ref{assumption3})&0.009 & 
$4.5 \times 10^{-8}$&$5 \times 10^{-12}$\\ \hline
Case III: (\ref{assumption4})& 0.002& 
$1.2 \times 10^{-25}$&$1.2 \times 10^{-11}$\\ 
\end{tabular}
\end{ruledtabular}
\end{table}

Notice that in all of the cases listed above the bound on 
$B(\tau \rightarrow e\, \gamma) <1.1 \times 10^{-7}$ \cite{bbar2} which is weaker than the other
two is trivially satisfied. That is the primary reason for using
the constraints from the LFV processes $\mu \rightarrow e\, \gamma$ and
$\tau \rightarrow \mu\, \gamma$.

In summary, Case II appears to be the most interesting one in that
both LFV branching ratios could in principle be observed.
This scenario has also another interesting phenomenological consequence:
The primary decay mode of a given mirror charged lepton is into a muon instead of
a tau as in the first scenario or an electron as in the third scenario. What would a typical decay length be? 

In the simple scenario described above, the primary 
decay mode of the mirror charged leptons 
is $\mu + \phi_S$ (assuming e.g. that $\nu^{M}_R$ is approximately
degenerate with its charged counterpart) with a coupling of the order $< 9 \times 10^{-3}$. For example, the decay rate of
$e^{M}_3$ is approximately $\Gamma(e^{M}_3 \rightarrow \mu + \phi_S) \sim m_E \lambda^2/(32\,\pi)$
and a decay length $l =1/\Gamma(e^{M}_3 \rightarrow \mu + \phi_S)$. With the bound on
$\lambda < 9 \times 10^{-3}$, one estimates the decay length to be $l > 2445\, fm$, which is {\em microscopic}.
A {\em macroscopic} decay length of the order of a few centimeters would imply a {\em much smaller}
$\lambda$ rendering the LFV processes discussed here practically {\em unobservable}.

\section{Conclusion}

The electroweak-scale right-handed neutrino model \cite{hung1} has a number of phenomenological
implications which could be explored experimentally in the near future. As mentioned
in \cite{hung1}, the see-saw mechanism could be directly tested at colliders by searching
for like-sign dilepton events coming from the production and decays of a pair of right-handed neutrinos
into a pair of like-sign lighter charged mirror leptons. The subsequent decays
of those charged mirror leptons into SM leptons, $e^{M}_R \rightarrow e_L + \phi_S$ ($e^{M}_R$
and $e_L$ are generic notations), provide the desired signals. How far from the beam pipe
these decays occur will depend on the strength of the Yukawa interactions written down
in Eqs. (\ref{yul1p}, \ref{yuk2p}). These Yukawa interactions are found to be proportional
to the Dirac neutrino mass matrix which enters the see-saw mechanism and vanish in the limit
where the light neutrino mass goes to zero.
It turns out that these interactions also generate
at one-loop level LFV processes such as the ones discussed here, namely
$\mu \rightarrow e\, \gamma$ and $\tau \rightarrow \mu\, \gamma$. (The subject
of $\mu-e$ conversion will be treated elsewhere.)
These LFV processes
put some interesting constraints on the model. In one example, it is found
that if one process is observed (which by itself is already astonishing), the other
will not be far from the current bound. Furthermore, within the framework
of this model, a ``macroscopic'' decay length (a cm or so) of the mirror charged lepton,
if discovered, implies that these LFV processes would be practically unobservable.
Conversely, if any of these LFV processes is observed, the decay length would
be tiny. This would require an extremely careful analysis to distinguish
these like-sign dilepton events from possible backgrounds.

In summary, our model contains several phenomena which can be tested at future
experimental facilities: 1) LFV processes at MEG and/or B factories; 2) Electroweak-scale
right-handed neutrinos at future colliders (LHC, ILC).

\begin{acknowledgments}
I would like to thank the Aspen Center for Physics
where this work was initiated and the LNF and Fermilab theory groups for hospitality
where part of this work was carried out.
This work is supported in parts by the US Department
of Energy under grant No. DE-A505-89ER40518. 
\end{acknowledgments}

\end{document}